**Simulating outcomes of interventions using a multipurpose simulation program based on the Evolutionary Causal Matrices and Markov Chain**


Hyemin Han[1], Kangwook Lee[2], Firat Soylu[1]

[1] Educational Psychology Program, University of Alabama, Tuscaloosa, AL 35406, USA

[2] School of Electrical Engineering, Korea Advanced Institute of Science and Technology, Daejeon 34141, Republic of Korea

Corresponding author:

Hyemin Han

Phone: 1-205-348-0746

Email address: hyemin.han@ua.edu




Simulating outcomes of interventions using a multipurpose simulation program based on the

Evolutionary Causal Matrices and Markov Chain

## Abstract

Predicting long-term outcomes of interventions is necessary for educational and social policy-making processes that might widely influence our society for the long-term. However, performing such predictions based on data from large-scale experiments might be challenging due to the lack of time and resources. In order to address this issue, computer simulations based on Evolutionary Causal Matrices and Markov Chain can be used to predict long-term outcomes with relatively small-scale lab data. In this paper, we introduce Python classes implementing a computer simulation model and presented some pilots implementations demonstrating how the model can be utilized for predicting outcomes of diverse interventions. We also introduce the *class-structured simulation module* both with real experimental data and with hypothetical data formulated based on social psychological theories. Classes developed and tested in the present study provide researchers and practitioners with a feasible and practical method to simulate intervention outcomes prospectively.

Kyle Reese: And it was over. Skynet was gone. And now one road has become many. Though questions remain, we'll search for the answers together. But one thing we know for sure. The future is not set.

-   *Terminator Genisys*



## 1. Introduction

Given the considerable resource demands of implementing large-scale and long-term interventions in authentic settings, predictions on expected outcomes of educational and psychological interventions based on small-scale studies are highly useful. The previous intervention studies have demonstrated that even one-time, short-term implementation of interventions can produce significant long-term effects; for example, short-term interventions promoting students' motivation and social adjustment were shown to have lasting effects even after a couple of years [1–3]. Thus, policymakers and educators should carefully consider such long-term, large-scale impacts of interventions when they intend to apply newly developed intervention models. However, it would be difficult to predict such long-term, large-scale outcomes solely based on lab data, collected during a short-term, from a small population. Furthermore, conducting large-scale intervention experiments to examine long-term, large-scale outcomes would be difficult due to the costs and insufficient resources [4].

Hence, we consider the computer simulation method as a potential candidate to address this issue. In the present study, we aimed to develop multipurpose, flexible Python classes for simulating outcomes of interventions based on relatively small data sets, such as data from experimental lab studies. This simulation model was designed to enable users to simulate predicted outcomes in diverse intervention conditions through iterative evolution processes. This model was founded by the ideas of evolutionary causal matrices (ECM) and Markov chain [1, 5].

### 1.1. Basics and Theoretical Foundations

#### 1.1.1. Evolutionary Causal Matrices (ECM)

 Evolutionary biology developed ways of modeling how well individuals with specific traits in a population can survive and reproduce successfully in an environment with pre-



determined selective pressures. These efforts have enabled us to predict states of equilibrium in the long run under a certain selection pressure [5, 6]. We can utilize such theoretical framework to examine effects of socio-cultural factors as well as natural factors. The changes in populations in a specific system influenced by such socio-cultural factors can be predicted by simulating evolutionary trends over time in the system.

For example, a previous study simulated how different type of cultural norms influence people's social behavior in terms of their norm conformity based on the theoretical framework of cultural evolution [1, 5, 7]. In this simulation study, researchers set two different types of populations, conformers and non-conformers. Conformers are people who tend to conform to existing social norms, while non-conformers do not. They also set two different types of hypothetical cultural systems. The first cultural system ($C_1$) has sufficient resources that can be distributed to all individuals. In C1, we can expect that norm conformers get more benefits compared with non-conformers who might spend additional energy to violate the currently available norms and rules. Consequently, this system will have more norm conformers, and non-conformers will change into conformers over time. On the other hand, the second system (C2) does not have sufficient resources available to individuals. In this case, non-conformers are more likely to be successful. As a result, the number of non-conformers will increase, and conformers will change into non-conformers over time in C2. The relative ratio of each type of individuals in each system will arrive eventually at an equilibrium as proposed by the theory of evolution.

If we can count the number of individuals in each status, such as conformers and non-conformers in the previous example, in a specific system at a specific time, $t$, then we can also quantify transitions between different statuses over time in the system. For example, in $C_1$, 80% of current conformers remain the current status a year later, while 20% of them become non-



conformers. Meanwhile, 60% of current non-conformers change into conformers a year later, while 40% of them remain in the same status. On the other hand, in $C_2$, 90% of current non-conformers maintain their current status a year later, while 10% of them become conformers. Meanwhile, 70% current conformers become non-conformers a year after, while 30% of them maintain their current status. In $C_1$, the number of conformers in the next year can be estimated by calculating 80% × number of current conformers + 60% × number of current non-conformers, and that of non-conformers can be estimated by calculating 20% × number of current conformers + 40% × number of current non-conformers. Following the same way, in $C_2$, the number of conformers in the next year becomes 30% × number of current conformers + 10% × number of current non-conformers, and that of non-conformers becomes 70% × number of current conformers + 10% × number of current non-conformers. If we repeat the same calculations, we can predict the long-term changes in systems quantitatively. We can express mathematically the longitudinal transitions between statuses in systems in the form of ECM for long-term predictions and simulations [5].

In general, the ECM are a series of matrices representing transitions between $t$ (indicating a specific time point) and $t+1$ (the next time point in time-series data) in a certain evolutionary system [6, 7]. A certain matrix constituting ECM demonstrates the changes in the numbers of populations situated in certain statuses between $t$ and $t+1$. For instance, consider an evolutionary system that can be explained by a sample matrix, a certain constituent matrix in ECM, represented in Table 1. If we say that 100 subjects are situated in each of three conditions (A, B, and C) at $t_0$, then the number of subjects in each condition at $t_0+1$ becomes 130 (= 100×(.70 + .50 + .10)), 70 (= 100×(.20 + .30 + .20)), and 100 (= 100×(.10 + .20 + .70)), respectively.



Thus, denoting the fraction of subjects in status *S* at time t by $F_S(t)$, the following recursive equation holds [1, 7].

*<Place Table 1 about here>*

$$F_S(t + 1) = \frac{\sum_{i \in P} F_i(t) ECM_{iS}}{\sum_{i \in P}\left[F_i(t) \sum_{j \in p} ECM_{ij}\right]}$$

$ECM_{ij}$ refers to the element in $i^{th}$ row and $j^{th}$ column of the ECM, which represents the transition rate between condition *j* at *t* to *i* at *t* + 1.

### 1.1.2. Markov Chain

In this section, we briefly overview the concept of Markov chain, a mathematical tool that is closely related to the ECM. Markov chain is primarily used to model and analyze systems whose states evolve randomly over time and has been extensively studied and applied in a variety of scientific and engineering fields [8]. For instance, Google's PageRank [9] algorithm, which is used to rank popular webpages, is based on a Markov chain model of user behavior. Another useful application of Markov chain is Markov chain Monte Carlo (MCMC) methods. Instead of capturing the dynamic of stochastic systems via the Markov chain, the MCMC methods leverage the mathematical property of Markov chains to approximately compute some quantities that cannot be computed directly or to obtain random sample from a complicated distribution that does not allow for directly sampling.

Before we present a formal definition of Markov chains, we first define some notation. Denote the *state* of a Markov chain at time t by $X(t) \in S$, where *S* is the (finite) state space. Note that the state space *S* varies across applications. For instance, $S = \{0, 1, 2, \dots\}$ if one wants to capture the amounts of dollars that a poker player has at time t, $S = \mathbb{R}^3$ if one wants to capture the 3D coordinates of a flying drone, and $S = \{0,1\}$ if one wants to capture a binary state. A Markov-chain is called *discrete-time Markov Chain* if its time-axis is discretized, i.e., t $\in$



$\{0, 1, 2, \dots\}$, and the state of the Markov chain changes only at those discretized time indices. If the time-axis is continuous, and the state of the Markov chain changes at any time, it is called continuous-time Markov Chain. In this work, it is sufficient to limit our scope to discrete-time Markov chains.

A discrete-time Markov Chain is a stochastic process that satisfies the following 'memoryless' property.

$$P(X(t+1) = x_{t+1} | X(t) = x_t, X(t-1) = x_{t-1}, \dots, X(0) = x_0) = P(X(t+1) = x_{t+1} | X(t) = x_t)$$

That is, regardless of how the system reaches the state $x_t$ at time $t$, i.e., the history of the dynamic system before time $t$, the transition from time $t$ to time $t+1$ (or any other future states of the system) depends only on the state $x_t$ at time $t$. Thus, one needs to specify the transition behaviors between consecutive time slots to fully characterize a Markov chain. For instance, the transition between time 0 and time 1 can be fully specified by a matrix $P^{(0)} \in [0,1]^{|S| \times |S|}$ whose (i,j) element $P_{i,j}^{(0)} = P(X(1) = j | X(0) = i)$. Note that the sum of each row is 1 to have a valid probability distribution. Such a matrix is called a *transition matrix*. In a similar way, one can specify transition matrices for every time slot, say $(P^{(0)}, P^{(1)}, \dots)$ to fully characterize a Markov chain. A Markov-chain is called time-homogeneous if $P^{(t)} = P$ for all t, and is called time-inhomogeneous otherwise.

One can observe that time-inhomogeneous Markov chains can precisely capture the idea of ECM. That is, one can appropriately choose transition matrices based on the given ECMs and one can precisely model the dynamics of the system. Hence, by leveraging a variety of tools developed for Markov chains to analyze the ECM, we develop tools for running Monte Carlo simulations (which estimate numerical outcomes for large-scale phenomena through repetitive random sampling) of Markov chains, and hence systems that can be modeled by ECM.



## 1.2. Current Study

Use of Markov chains to predict outcomes of interventions is part of a larger trend, for example in psychological sciences, where prediction as opposed to mere explanation takes precedence in evaluating how well a theory can account for an phenomenon studied [10]. The distinction between explanation and prediction mainly concerns how a causal model is evaluated. An explanatory model is tested based on how well it can account for the relations between independent and dependent variables in data collected from a representative sample. Explanatory models have three main weaknesses. One is *overfitting*, incorporating noise or causal relations specific to the sample (but not to the population or, more widely, the phenomenon studied) into the model. Due to the problem of overfitting the model that best explains the data is not always the one that best predicts future behavior [11]. The second is *p*-hacking [12] – also called data dredging or data snooping; finding statistically significant patterns that are not part of the a priori hypotheses. In this case, the casual relations emerging from data mining are not driven by a priori theorizing. Given a large data set and the wide range of data mining methods one can almost surely find significant effects in any data set, but again these effects actually can be due to sample-specific noise and are not generalizable, increasing false-positive rates. The problem of overfitting can be addressed by *cross-validation*; which involves building a model with one sample-specific data set and testing it with a data set from a different, independent, sample. In this way casual relations presented in the model that are actually due to sample-specific noise will be less likely to be validated with the data from a different sample. Therefore, cross-validation provides a better way of estimating how well a model can generalize to new data. An additional way of improving generalizability is *regularization*, constraining the model with prior



knowledge (a priori theorizing). In this way statistical analysis is not a "fishing trip" for significant effects but a test of a priori and theory-driven models and hypotheses.

Problems of overfitting and p-hacking have lead to what is now called a "replication crisis" most notably in psychology and medicine, where findings across many studies show diverging results and are not replicated [13]. Some of the solutions proposed to remediate the underlying causes for this crisis included more stringent protocols and conventions for research [12], Bayesian statistics and meta-analysis [14], more appropriate and balanced use of explanatory and predictive modeling [11], and use of predictive machine learning methods [10]. Use of predictive machine learning methods provide enhanced ways of responding to concerns with p-hacking and overfitting. Cross-validation is an inherent aspect of using machine learning for prediction. Once the predictive model is trained with one data set, it has to be tested with a separate set to measure its predictive prowess. Regularization is achieved by using cost functions that more stringently penalize deviations from the predicted model, yielding simpler and more generalizable models. Use of predictive machine learning also decreases the likelihood of introducing an intentional bias in the statistical analysis, for example opting for conducting inferential statistics for specific patterns based on visual inspection of data, even though these comparisons may not be part of the a priori planned analysis. Finally, predictive models are more easily used in applied settings, narrowing the gap between research and practice [11].

The previous studies on predictive modeling, for example in psychology, have demonstrated that predictive simulation models produce reliable and valid outcomes coherent with theoretical assumptions [1, 15]. However, the previous simulation models were not able to simulate complicated cases properly. For instance, Han et al.'s (2016) simulation program written in MATLAB was initially designed to predict long-term outcomes of educational



interventions when only one type of intervention was applied with a regular interval. Hence, in order to address this limitation and make the simulation model applicable to diverse situations, the present study aims to develop a general-purpose simulation program enabling users to predict outcomes in more complicated cases, such as when more than one type of interventions is conducted, and when the intervals between occurrences of interventions are irregular.

We developed an enhanced simulation program, in Python, based on ECM and Markov chain for prospective uses in the fields of education and psychology. In order to test this program, we conducted four simulations predicting long-term outcomes of interventions with diverse conditions and intervals. For those simulations, we created ECM based on findings from the previously conducted moral educational intervention experiments [16–18] (for simulations 1 and 2) and hypothetical ECM based on the previous social psychological experiments (for simulations 3 and 4). In addition, we performed statistical analyses, including t-tests, mixed-effects analyses, and ANOVA, comparing outcomes between different intervention conditions and different intervention intervals in order to test whether our simulation program was able to produce simulation outcomes coherent with the previous intervention experiments and psychological theories for explanatory purposes.

## 2. Materials and Methods

### 2.1. Problem formulation and algorithm

We defined following variables for the explanation of our algorithm:

groups: Number of different population (or sample) groups. For example, in the case of our educational intervention simulation, groups = 2 (1: non-participants, 2: voluntary service participants). In terms of the Markov chain terminology, this corresponds to the cardinality of the state space $|S|$.



length: The total length of simulation. It indicates how many iterations will be performed. We set length = 100 for our simulation. In terms of the Markov chain terminology, this is the number of time steps that are simulated.

t: Current time point. t ranges from 1 to length. At the end of each iteration, t increases by 1.

conditions: Number of different types of interventions simulated. In terms of the Markov chain terminology, this indicates the number of different Markov chains we simulate.

status_t_now (t = 1 to length, group = 1 to groups): Number of populations (or samples) situated in a specific t in each group (group). For example, in the case of our simulation, the value in status_t_now (1,1) indicates the number of non-participants before the beginning of the simulation.

schedule (t = 1 to length): Intervention schedule. A stored value indicates which type of intervention is performed at a specific t. In terms of the Markov chain terminology, this indicates the times at which certain transition matrices are applied to the Markov chain.

ECM (condition = 1 to conditions, group = 1 to groups): Evolutionary causality matrix for each intervention condition. status_t_now is being updated during each iteration by calculating the dot product of status_t_now and ECM (schedule (t), *). In terms of the Markov chain terminology, this is the transition matrix corresponding to each intervention condition.

We aimed to study how to predict the long-term outcomes of educational interventions efficiently when interventions types (schedule (length)), ECM for each intervention type (ECM (conditions, groups)), and initial states were given (status_t_now (1, groups)). Thus, our simulation algorithm was developed to estimate the number of populations (or samples) in each group at specific t (status_t_now) by calculating the dot product of the current status



(status_t_now) and ECM (ECM) iteratively. It implemented the way of transition calculation explained in the introduction section. In order to examine the effects of different types of interventions, multiple ECM were created; a matrix was created for each designated intervention condition. At a specific t, the type of matrix that was used for the dot production was determined by a pre-set intervention schedule (schedule).

To evaluate the quality of our simulations, we compared the simulated outcomes with actual intervention experimental data collected from classrooms [18]. Moreover, we also compared whether the simulate outcomes were consistent with what could be predicted from developmental and social psychological theories related to the ideas of zone proximal development (ZPD) and attainable goal setting [19, 20]. We compared the number of student populations situated in different groups (i.e., non-participants and service participants) after conducting different types of educational interventions (i.e., attainable versus unattainable exemplar story interventions, non-moral interventions for the control condition) with different intervention schedules.

The core algorithm of this iterative calculation process is explained in a flowchart (see Fig. 1)

<div align="center"><em><Place Fig. 1 about here></em></div>

### 2.2. Intervention outcome simulation program

We created classes for the simulation of outcomes of interventions based on ECM and Markov chain. These classes were developed in Python. All Python source codes for classes and simulation tutorials are available online as supplementary materials. These are downloadable from the GitHub, https://github.com/xxelloss/Markov-Learning, or from the Dataverse, http://dx.doi.org/10.7910/DVN/KBHLOD [21]. Readers can learn how to create and conduct



their own simulation by following the tutorials downloadable from the aforementioned repository. There were two main classes constituting the simulation program: Markov_learning (Markov_learning.py) and ECM_matrix (ECM_matrix.py).

Markov_learning class deals with the simulation model creation and evolution processes. Users can define a specific simulation model using this class while setting multiple intervention schedules containing different intervention conditions for further statistical comparisons. This class implements the evolution of a certain system that is designated to perform a certain intervention schedule from $t$=0 to a specified period. ECM_matrix class contains a constituent matrix of ECM corresponding to a certain intervention condition. This class consists of a matrix representing transitions between $t$ and $t+1$ for each status. Methods included in this class support the matrix creation and iterative calculation.

In each simulation model, one Markov_learning class is created. Users need to provide required information, i.e., the number of intervention conditions (Markov_learning.conditions), the number of different statuses (Markov_learning.size), the length of evolution period (from $t =$ 0 to the end) (Markov_learning.length), and the number of intervention schedules to be analyzed (Markov_learning.schedules), as parameters at the moment of the creation of this class. Once a Markov_learning class is created with all the required parameters, it automatically creates multiple ECM_matrix classes (Markov_learning.ECM) according to the preset number of intervention conditions; each ECM_matrix class is designed to be corresponding to each intervention condition. Users then set each constituent matrix representing transitions between time points and statuses for each intervention condition that is contained by an ECM_matrix class (ECM_matrix.matrix).



Before conducting iterative calculation processes for evolution, users are required to set initial statuses at $t$=0 (Markov_learning.status_t0) and intervention schedules (Markov_learning.schedule). First, initial statuses at $t$=0 should be set. The number of statuses follows a pre-set number, Markov_learning.size. For example, in the sample case presented in the introduction section, the initial statuses can be set in the form of an array, [100, 100, 100]. Second, intervention schedules to be compared should also be determined. For instance, once we set two different intervention conditions, A and B, we can set multiple intervention schedules that will be compared after the end of the evolution process. If N represents the absence of intervention (control condition), then we may set several intervention schedules, such as:

Schedule 1: A, N, A, N, A, N, A, …, N, A, N

Schedule 2: A, N, N, A, N, N, A, …, A, N, N

Schedule 3: A, B, N, A, B, N, A, …, A, B, N

Schedule 4: A, B, N, A, B, N, N, …, N, N, N

…

Schedule 10: N, N, N, N, N, N, N, …, N, N, N

The length of each schedule should be identical to the length of intervention periods, specified during the creation of the Markov_learning class (Markov_learning.length). Also, the number of schedules to be determined should be identical to the preset number of schedules in the class, Markov_learning.schedules. Unlike the case of the previous simulation program [1] that only enabled users to set intervention schedules containing only one intervention condition and control condition with regular intervals (e.g., Schedules 1, 2, and 10), the simulation program developed in the present study enables users to set schedules with multiple intervention conditions with irregular intervals (e.g., Schedule 3 and 4).



Once all intervention conditions and schedules are set, evolution processes that are constituted by a series of iterative transitions between $t$ and $t+1$, from $t=0$ to Markov_learning.length, can be initiated. Transitions in statuses between $t$ and $t+1$ are calculated using the equation presented in the introduction section. If the current $t=0$, statuses at $t=1$ are calculated based on the preset Markov_learning.status_t0 values. Calculated statuses at $t+1$ are stored and updated in an array, Markov_learning.status_t_now. After the end of the whole evolution processes until $t$=Markov_learning.length, status values from $t=0$ to $t$=Markov_learning.length are printed out in text files for further statistical analyses. Each text file contains status values for each intervention schedule. The text files were analyzed with STATA.

### 2.3. Sample simulations and dataset

In the present study, we performed a total of five simulations as pilots using the developed classes. Three of them were based on an experimental dataset collected from intervention experiments that were previously conducted [16, 17, 22]. These three simulation trials were performed in order to predict long-term, large-scale outcomes of the educational interventions designed by the previous intervention experiments that were conducted within a small-scale context (labs and classrooms) during a short period (2-3 months). ECM for these three simulation trials were created from the real experimental data. Two other simulations were performed using ECM created from hypothetical cases. These simulation trials predicted outcomes from hypothetical intervention designs based on social and educational psychological theory [20, 23–26]. Further details pertaining to the nature of intervention experiments and datasets are elaborated in each subsection in the results section, focusing on each simulation trial.



### 2.4. Statistical analyses

We conducted several statistical analyses in order to test whether the developed simulation program could provide simulation results that were consistent with the findings from the real experiments (simulations 1 and 2) and psychological theories (simulations 3 and 4). In the case of simulations 1 and 2, we utilized t-test and mixed-effects analysis methods to test whether the different types and frequencies of interventions produce significantly different outcomes as shown by the previous social psychological experiments [16, 17]. In the cases of simulations 3 and 4, we conducted mixed effects analyses to examine whether different types of hypothetical interventions might produce significantly different longitudinal outcomes as predicted based on the previous social psychological theories [20, 24, 25]. Further details pertaining to simulation and statistical methods are explained in each section below introducing each simulation trial.

## 3. Simulations and Results

### 3.1. Simulation 1

In simulation 1, we simulated the long-term outcomes of moral educational interventions based on experimental data collected by the previous experiments [16, 17]. These experiments tested what types of moral stories can effectively promote students' engagement in prosocial activities, particularly voluntary service activities. In these experiments, data collected from 54 college students was used. The students were randomly assigned to three groups: the attainable and extraordinary moral story groups, and the control group; 17, 18, and 19 students were assigned to each group, respectively. The attainable group was presented with moral stories demonstrating attainable voluntary service engagement (requires a commitment for less than two hours per week) written by peer college students. The students assigned to the extraordinary



group were presented with stories of service engagement that were perceived to be very difficult to emulate for college students (requires a commitment for more than ten hours per week). The control group was presented with non-moral stories, such as sports news reports. Before the beginning of the intervention session, each student was provided with a survey form asking the student's initial voluntary service engagement. During the intervention session, students were presented with the type of stories matching with their group assignment. 1.5 months after the end of the intervention session, a, post-test, voluntary service engagement survey form was distributed to the students.

From the collected dataset, we created three ECM for three experimental conditions. These ECM demonstrate changes in the number of service participants and non-participants between the pre- and post-test periods. Table 2 shows ECM for the attainable and extraordinary moral story conditions, and the control condition.

<p style="text-align:center"><em><Place Table 2 about here></em></p>

We conducted iterative evolution processes from $t$=0 to 100 (100 iterations) using the created ECM. The time gap between $t$ and $t$+1 was 1.5 months, which was identical to the actual time gap between the pre- and post-tests in the intervention experiment. We set [127 (participants), 111 (non-participants)] as the initial value at $t$=0. While conducting the evolution processes, we applied different intervention frequencies to examine what is the minimum required frequency of interventions to produce significant outcomes. More specifically, we set 50 different schedules with different intervals as follows:

Schedule 1 (no interval, one I per every 1.5 months): I, I, I, I, …., I, I, I, I

Schedule 2 (interval = 1, one I per every 3 months): I, C, I, C, …., I, C, I, C

Schedule 3 (interval = 2, one I per every 4.5 months): I, C, C, I, …, I, C, C, I



…

Schedule 50 (interval = 49, one I per every 75 months): I, C, C, C, …, C, C, C, C

(I: intervention, C: control condition)

We performed evolution processes for these 50 intervention schedules for both intervention types, attainable and extraordinary moral story interventions. After performing all evolution processes, we compared the mean number of service participants versus that of non-participants in each intervention schedule using the t-test. We examined the significance of the difference and resultant Cohen's $D$ value as the indicator for effect size in each intervention schedule. In the case of the calculation of $p$-values, we applied Bonferroni's correction for multiple comparisons, since we simulated three different experimental conditions.

*<Place Figs. 2 and 3 about here (Color)>*

Fig. 2 shows how the mean number of participants and non-participants changes according to the type and frequency of intervention. Fig. S1 demonstrates changes in the $p$-value, and Fig. 3 demonstrates Cohen's $D$ value from schedule 1 to 50 when the attainable moral story intervention was applied. Overall, the attainable moral story intervention positively contributed to the increase of service participants, and this result was in line with the original experimental study [17]. According to the result, the attainable moral story intervention should be applied at least once per every 36 months (schedule 24) in order to produce a statistically significant difference ($p < .05$) between the numbers of service participants and non-participants. Furthermore, in terms of the effect size, the intervention should be conducted at least once per every 15 months (schedule 10) to produce a large effect size ($D > .8$) or per every 36 months (schedule 24) to produce a medium effect size ($D > .5$).

*<Place Fig. 4 about here (Color)>*



Figs. 4 and S2 show outcomes when the extraordinary moral story and control conditions were compared. Overall, this intervention decreased the number of participants. The t-tests indicated that first, when this intervention was applied at least once per every 48 months (schedule 32), the difference between the numbers of service participants and non-participants became statistically significant at $p < .05$ (see Fig. S2). Second, the intervention should be conducted at least once per every 15 months (schedule 10) to produce a large effect size ($D > .8$) or per every 36 months (schedule 24) to produce a medium effect size ($D > .5$) (see Fig. 3).

Simulation 1 demonstrated how the developed simulation program can predict long-term outcomes of different types of interventions based on relatively short-term, small-scale experimental data. The overall simulation results were coherent with the findings from the previous intervention experiment that reported significant differences in the promotion of service participants between different intervention conditions. The simulation was also able to estimate the required minimum frequency of the application of interventions in order to produce significant outcomes with a large effect size.

### 3.2. Simulation 2

In simulation 2, we examined cases when interventions were applied irregularly. In other words, simulated intervention schedules in this simulation were designed to have irregular intervention intervals. We set 50 different intervention schedules (from $t$=0 to 100, 100 iterations) using the same dataset and ECM used in simulation 1. Also, identical to simulation 1, we used [127 (participants), 111 (non-participants)] as the initial value at $t$=0. From schedule 1 to 50, the intervals between interventions became sparser. Below are some examples for the schedules:

Schedule 1: I, C, I, C, I, C, I, C, I, C, I, C, I, C, I, …

Schedule 2: I, C, I, C, C, I, C, C, I, C, C, C, …



Schedule 3: I, C, I, C, C, C, I, C, C, C, C, C, I, C, …

…

Schedule 50: I, C, I, C, C, C, C, C, C, C, C, C, C, …

(I: intervention, C: control condition; irregular intervals)

We examined whether the number of service participants decreased as the intervention applied less frequently (higher schedule number) when the attainable story intervention applied by conducting mixed-effects analysis. The number of service participants was set as the dependent variable. The designated fixed effect was the schedule number, which was negatively associated with the intervention frequency. We set $t$ as the random effect.

<Place Fig. 5 about here (Color)>

Fig. 5 demonstrates change in the mean number of service participants and non-participants from schedule 1 to 50. The result of the mixed-effects analysis reported that the fixed effect, the schedule number, was significant in the mixed-effects model, $B = -.14$, $SE = .01$, $z = -19.01$, $p < .001$, 95% CI [-.15 -.12], $f^2 = .08$. In other words, service participation was less likely to be promoted as the attainable story intervention was applied less frequently. Unlike the previously developed simulation program that only allowed the simulation of intervention outcomes when interventions were applied regularly, simulation 2 showed that the simulation program developed in the present study was able to simulate cases with irregular application of interventions.

### 3.3. Simulation 3

In this simulation, we created a simulation model based on the theoretical framework of social psychology pertaining to how attainable and extraordinary exemplary stories differentially influence subjects' motivation. According to Bandura & Schunk's (1981) seminal work, setting



proximal goals, instead of distal goals, promotes self-efficacy, academic motivation, and

achievement more effectively [27]. Similarly, following experimental studies have shown that

presenting attainable exemplars and goals was more likely to promote motivation compared to

presenting extraordinary exemplars and goals [25, 26]. Merely presenting extraordinary

exemplars to people, particularly who did not originally participate in the same behaviors

presented by the exemplars, might backfire and decrease  the motivation for emulation [28, 29].

Instead, such extraordinary exemplars that are perceived not to be emulatable to ordinary people

might promote motivation among people who already engage in the behaviors of the exemplars.

Given the idea of ZPD proposed by Vygotsky (1978), such extraordinary exemplars might work

as scaffoldings for current participants; after watching more demanding exemplary stories, such

current participants might be motivated to do more challenging tasks.

Based on these psychological studies, we hypothesized that first, the presentation of

extraordinary exemplars might reinforce engagement among current participants, but might not

make non-participants start engagement. Second, attainable stories might effectively motivate

non-participants to initiate participation, but might not intensify the strength of engagement

among current participants. Thus, we created three hypothetical ECM, one for the attainable

story condition, one for the extraordinary condition, and one for the control condition (see Table

3). The nature of each condition was similar to that in simulation 1. Because the previous two

simulations based on real data did not differentiated different degrees of participation, in these

ECM, we differentiated current participants into two categories: high and low participants.

*<Place Table 3 about here>*

Similar to simulation 1, we performed evolution processes from $t$=0 to 100 (100

iterations) and compared 50 different intervention schedules. We set [100 (high participants),



100 (low participants), 100 (non-participants)] as the initial values at $t$=0. In order to see how the extraordinary story intervention boosted high participants, we included both the extraordinary and attainable story interventions in the schedules with different intervals. The intervals between the applications of the extraordinary story intervention were adjusted. Here are the created intervention schedules:

Schedule 1 (no interval): E, E, E, E, …., E, E, E, E

Schedule 2 (interval = 1): E, A, E, A, …., E, A, E, A

Schedule 3 (interval = 2): E, A, A, E, …, E, A, A, E

…

Schedule 50 (interval = 49): E, A, A, A, …, A, A, A, A

(A: attainable intervention, E: extraordinary intervention)

As shown above, the extraordinary story intervention was performed less frequently as the schedule number increased. As we hypothesized, the mean number of high participants was expected to decrease and that of low participants was expected to increase as the schedule number increased. In order to test these hypotheses, two mixed-effects analyses were conducted. For both analyses, we set the schedule number as the fixed effect and $t$ as the random effect. In the first analysis, the number of high participants was set as the dependent variable, while the number of low participants was set as the dependent variable in the case of the second analysis.

*<Place Fig. 6 about here (Color)>*

The results of the mixed-effects analyses demonstrated that first, the schedule number was negatively associated with the number of high participants, $B$ = -.12, $SE$ = .01, $z$ = -.21.30, $p$ < .001, 95% CI [-.13 -11], $f^2$ = .09 (see Fig. 6). Second, when the number of low participants was set as the dependent variable, the schedule number was positively associated with this dependent



variable, $B = .38$, $SE = .01$, $z = 27.91$, $p < .001$, 95% CI [.35 .41], $f^2 = .16$ (see Fig. 5). As the

application of the extraordinary story intervention became more frequent, the number of high

participants significantly increased. Instead, when the extraordinary story intervention was

performed less frequently, in other words, the frequency of the attainable story intervention

increased, the number of low participants increased. These findings were coherent with our

hypotheses and the previous psychological studies that have discussed the differentiated

influences of attainable and extraordinary exemplars on motivation among different populations.

### 3.4. Simulation 4

In this simulation, we used the ECM created in simulation 3 in order to simulate whether

the order of different types of interventions significantly influenced outcomes (e.g., attainable—

extraordinary vs. extraordinary —attainable). Here are five intervention schedules simulated:

Schedule 1: A, E, C, A, E, C, …., C, A, E, C, A, E

Schedule 2: E, A, C, E, A, C, …., A, C, E, A, C, E

Schedule 3: A, A, C, A, A, C, …, A, C, A, A, C, A

Schedule 4: E, E, C, E, E, C, …, E, C, E, E, C, E

Schedule 5 (control condition): C, C, C, C, C, C, …, C, C, C, C, C, C

(A: attainable intervention, E: extraordinary intervention, C: control condition)

Similar to the previous simulation, we performed evolution processes from $t$=0 to 100

(100 iterations) with the initial values [1000 (high participants), 1000 (low participants), 1000

(non-participants)] at $t$=0.

According to the previous psychological studies that have demonstrated different

influences of different types of exemplars on motivation [19, 20, 25, 26], we hypothesized that

the number of high participants, participants showing strong engagement, would be maximized



when the attainable story intervention was performed followed by the extraordinary story intervention. In order to test this hypothesis, we conducted mixed-effects analyses. The number of intervention schedules was set as the fixed effect; to examine which intervention schedule performed best, we treated the schedule number as a categorical variable. We set $t$ as the random effect. Similar to the previous simulation, both the numbers of high and low participants were used as dependent variables. Furthermore, we performed one-way ANOVA to figure out which intervention schedule outperformed the other in terms of the numbers of high and low participants. For the post-hoc test, we utilized Scheffe's method.

*<Place Tables 4 and 5 about here>*

The results from the mixed-effects analyses are presented in Table 4. As shown, we found significant differences in the numbers of three different types of participants between different intervention schedules. Furthermore, the results of the one-way ANOVA and post-hoc tests also demonstrated significant differences as well (See Table 5). In the case of the high participant number as the dependent variable, schedules 1 and 4 showed the similarly highest, while the control condition, schedule 5, showed the lowest number. In the case of the number of low participants, schedules 1 and 2 both showed the greatest mean values, while schedules 4 and 5 both showed the smallest mean values. Finally, when we examined the number of non-participants, schedule 5 showed the highest value, while schedules 1 and 3 both showed the lowest values; in other words, the number of participants including both high and low participants was highest when schedule 1 or 3 was applied. Given these findings, in order to maximize participation regardless of its degree, schedule 1 or 3 was most effective. Between these two schedule, schedules 1 and 3, schedule 1 was more effective to increase the number of high participants. These findings supported our hypothesis based on the previous psychological



studies that the attainable story intervention followed by the extraordinary story intervention should be applied to effectively promote overall engagement, and eventually, strong engagement.

## 4. Discussion

Large-scale implementation efforts require making decisions about duration, frequency, and sequencing of different types interventions to maximize learning outcomes. To facilitate decision-making about these implementation parameters, we developed and tested a new multipurpose simulation program, implemented in Python, enabling researchers, educators, and policy makers to simulate long-term outcomes of interventions using relatively small data sets. Unlike the previously published studies that also used ECM and Markov chain to address a specific research question with a single dataset [1, 15], we developed a simulation program that can perform simulations with different datasets, including both a real lab dataset and a hypothetical dataset created based on the previous psychological studies. In addition, the presented program allows simulating multiple interventions with different duration and sequencing configurations. This builds on our previous work that focused on simulating outcomes of a single intervention with varying durations and frequencies [1]. Findings from the present study suggested that users will be able to utilize our simulation program to scale-up their intervention designs that were initially developed in a small scale or to set hypotheses for long-term, large-scale experiments based on small-size pilot data.

As we mentioned in the introduction, even brief psychological and educational interventions can produce significant long-term outcomes, so researchers, educators, and policy makers should be cautious when they intend to apply interventions initially developed in a lab to real educational settings. In order to address this issue, our simulation program can provide useful insights regarding types of interventions, frequencies of intervention application, and



intervention schedules in order to maximize positive intervention outcomes. Furthermore, as we did in simulations 3 and 4, it is possible to create hypothetical ECM that reflect theoretical assumptions and simulate long-term evolution processes with our program. Because it would be difficult to conduct large-scale longitudinal experiments to measure the actual effectiveness of interventions, due to limited time and resources, the developed simulation program will enable researchers to conduct such experiments, to formulate hypotheses for their experiments based on the previous studies. In addition, because we programmed Python classes to implement the simulations, instead of programming customized simulation programs, the present study can enable users to create their own simulation models using their own datasets. Users can perform their own simulations by modifying one of the tutorial source codes that implemented the four simulations.

Using computational models to support decision making about large-scale implementations aligns with the current emphasis on evidence-based practices in education [30], burgeoning efforts in connecting insights from lab studies with educational practice and policy-making [31], and proposals on enhanced rigor of predictive models over explanatory models in psychology [10]. Here we presented a platform for the application of Markov models to extrapolate results from short-term, small-scale intervention studies to inform large-scale implementations. Computational predictive models provide researchers and practitioners with an analytic method to connect evidence from smaller-scale lab or classroom studies with larger-scale practices and policies. One impediment for mainstream adoption of such methods is the technical complexities and difficulties of developing computational models for educational researchers, who are not trained with these methods. We addressed this by developing and making available a set of Python libraries. The simulation platform presented here can be used



by researchers doing work in a wide range of fields in education (e.g., special education, literacy, STEM education).

Predictions for future outcomes can also be conducted by using *deep learning* [1, 32]. Deep learning is a promising machine learning algorithm that can efficiently learn from collected data (or so-called training data) and expose implicit hidden patterns or structures within the data, called the *model*. Once the model is learned, it can be used to predict unobserved cases, often achieving the state-of-the-art prediction accuracies over any other machine learning algorithms. Various deep learning architectures for different target applications have been proposed and shown successful in the literature: CNN (Convolutional Neural Network) for computer vision tasks and RNN (Recurrent Neural Network) for tasks involving sequences (such as speech recognition and language processing) are the prototypical architectures [28-30]. Indeed, one may be able to achieve an improved prediction performance by applying deep learning, especially the RNN-type architecture, to our task.

However, we decided to use the ECM and Markov chain method in the present study given its strengths compared with the deep learning method. Although the deep learning method can detect and model complex nature of collected data, models learned by deep learning are constituted by numerous hidden layers with nonlinear activations, making them black-boxes that are not interpretable. Further, the deep learning approach typically requires a much larger amount of training data due to its large number of parameters, and require larger computational power [31, 32]. In contrast to these drawbacks of the deep learning models, our ECM-based approach provides interpretable model parameters and prediction results [6] as well as possesses much fewer number of parameters, resulting in low computational overheads [1].



There are several limitations in the present study. First, although we developed Python classes for simulations, users who do not have any knowledge of Python and computer programming cannot conduct their own simulations with our program. This limitation might be critical to educators and educational policy makers who are potential users of this program but without programming skills. In order to address this issue, a graphical user interface (GUI) providing a visualized access to the classes should be developed. We plan to address this limitation by developing a GUI module that provides end users with easy access to the developed classes. We will develop both a cross-browser GUI with PyJamas and class-platform GUI with appJar in order to enable end users to use these classes with ease, both in online and offline modes.

Second, this simulation module based on ECM and Markov chain can only predict outcomes in the form of categorical variables; it would be difficult to include covariates (e.g., demographical variables) other than designated ECM to the simulation model. These limitations might prevent potential users who intend to examine effects of associations and interactions between multiple variables. Hence, future studies should test other computer simulation methods, such as the aforementioned deep learning, in order to implement more sophisticated simulations for complicated cases. We intend to develop a deep-learning based simulation program with Google's multipurpose deep learning tool, TensorFlow. We are currently conducting a pilot test with TensorFlow to simulate the intervention dataset. Results from this deep learning simulation will be compared with those from traditional prediction methods, such as regression analysis, to examine whether the deep learning method can allow us to predict intervention outcome more accurately.



Third, the developed simulation can only be used for predicting outcomes and not for inferential statistics. Although we presented some results associated with statistical significances and effect sizes, these results were introduced only for explanatory purposes, not for suggesting that future users can use our simulation classes for statistical tests. We consider implementing the developed classes in R that provides functionalities for more sophisticated statistical analyses in order to address this limitation.

## 5. Supplementary Materials

All Python source codes for classes and simulation tutorials are downloadable from the GitHub, https://github.com/xxelloss/Markov-Learning, or from the Dataverse, http://dx.doi.org/10.7910/DVN/KBHLOD [21]. numpy, pandas, statsmodels, and scipy should be installed to use Markov-Learning.

**Figures and captions**

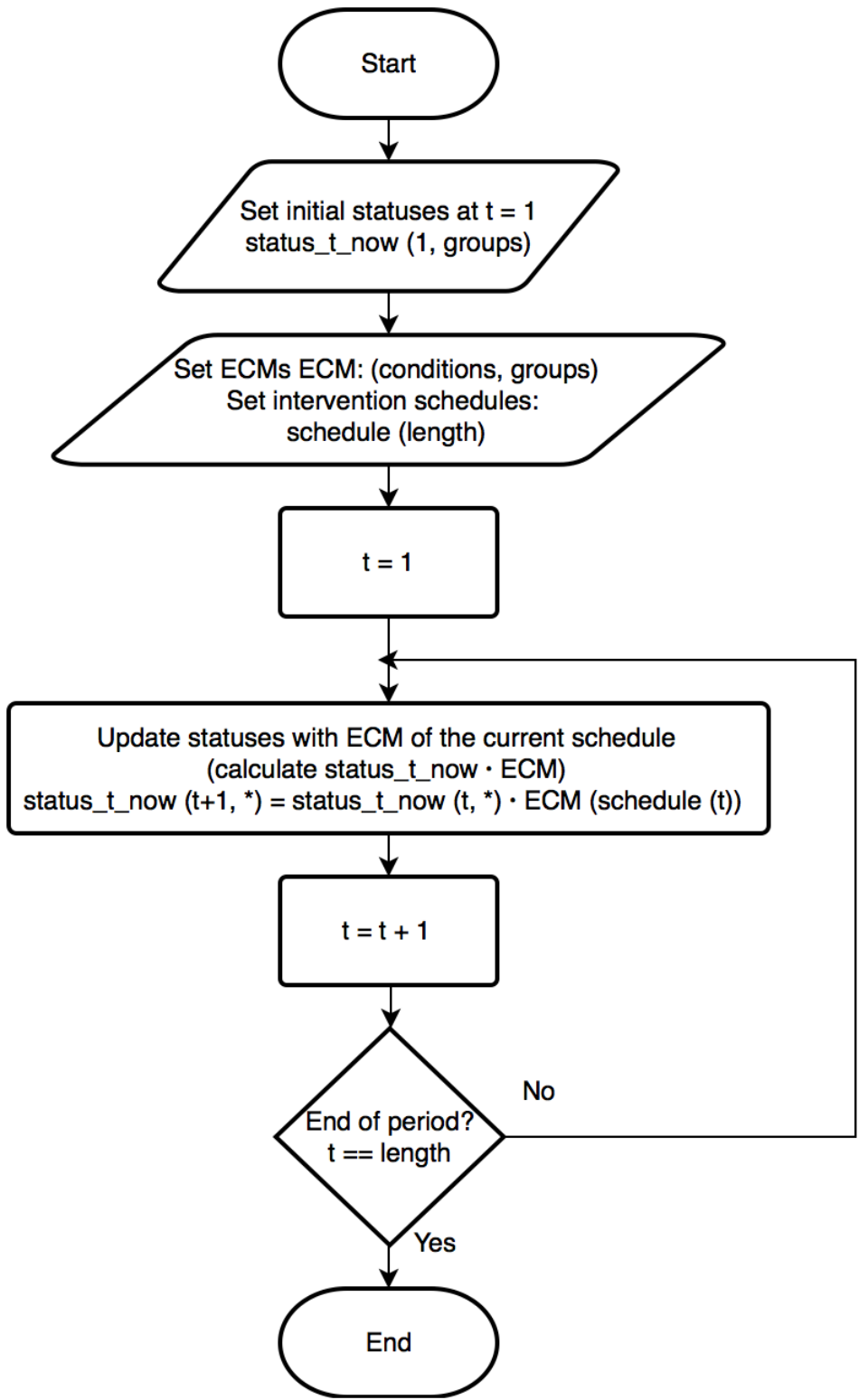

Fig. 1 Flowchart describing the core algorithm of our simulation model with ECM



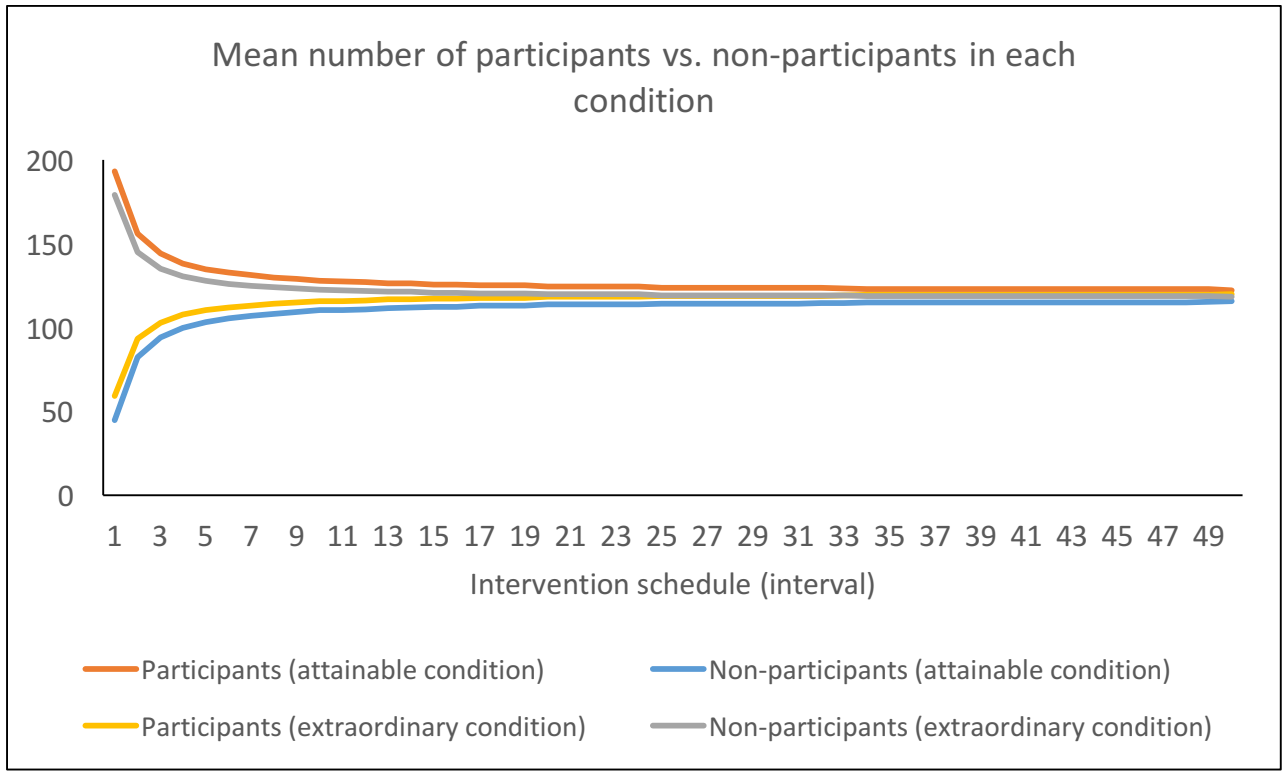

**Fig. 2** Change in the mean number of participants and non-participants in each condition across different intervention schedules (intervals) in simulation 1



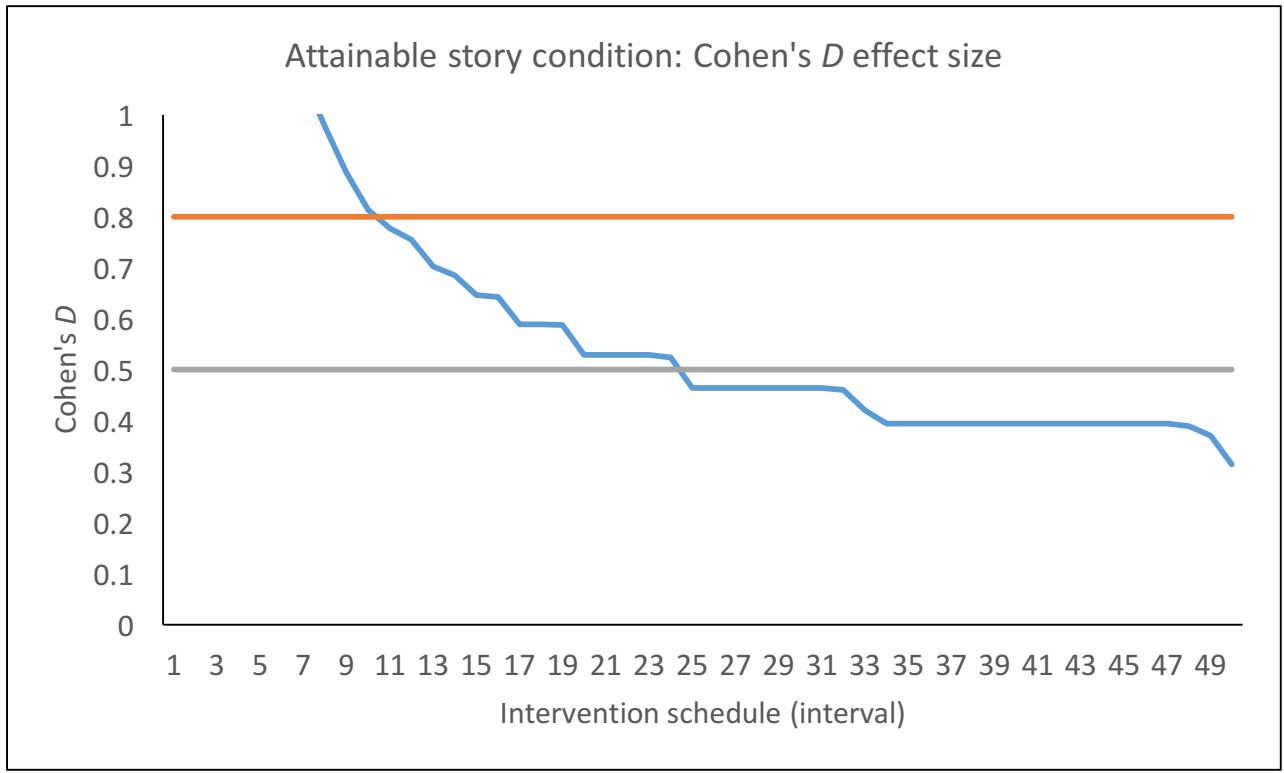

**Fig. 3** Cohen's *D* value in the case of the attainable story intervention condition in simulation 1. Blue line: Cohen's *D* value resulted from the t-test comparing participation rate between the attainable story intervention condition vs. control condition per different intervention intervals. Orange line: a threshold of *D* = .8 (large effect size). Gray line: a threshold of *D* = .5 (medium effect size).



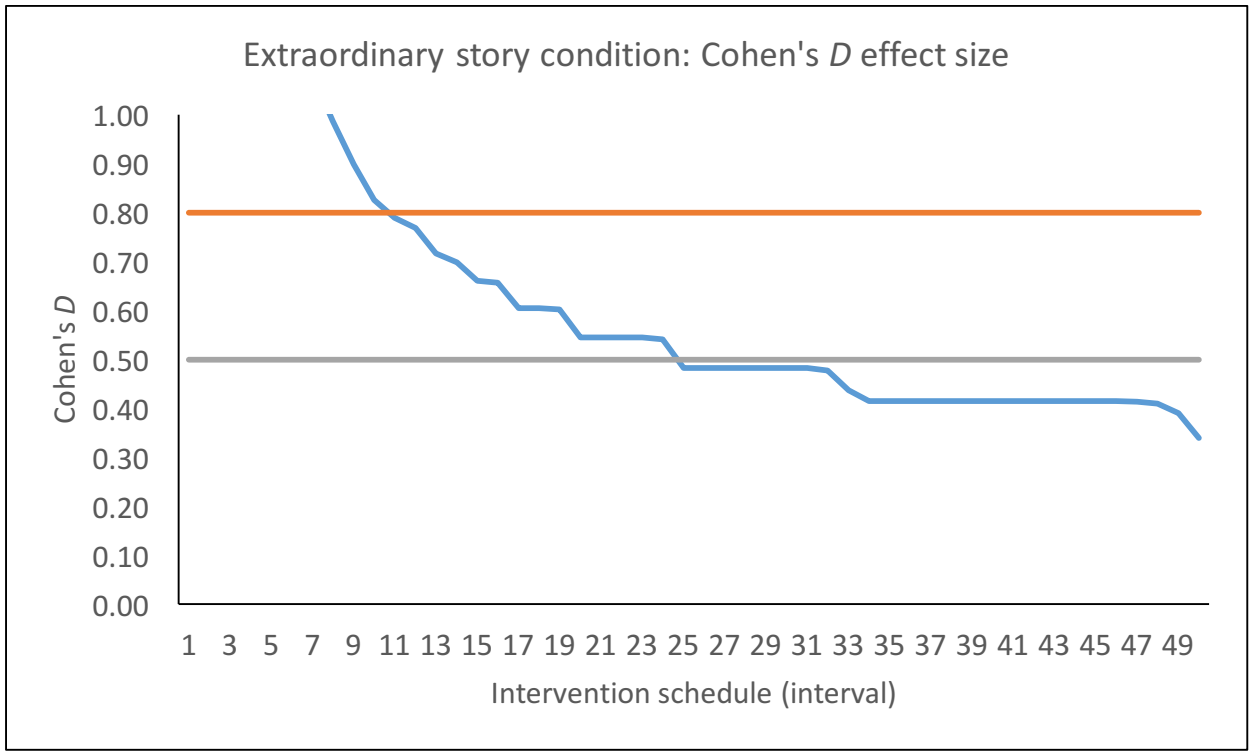

**Fig. 4** Cohen's *D* value in the case of the extraordinary story intervention condition in simulation 1. Blue line: Cohen's *D* value resulted from the t-test comparing participation rate between the extraordinary story intervention condition vs. control condition per different intervention intervals. Orange line: a threshold of *D* = .8 (large effect size). Gray line: a threshold of *D* = .5 (medium effect size).



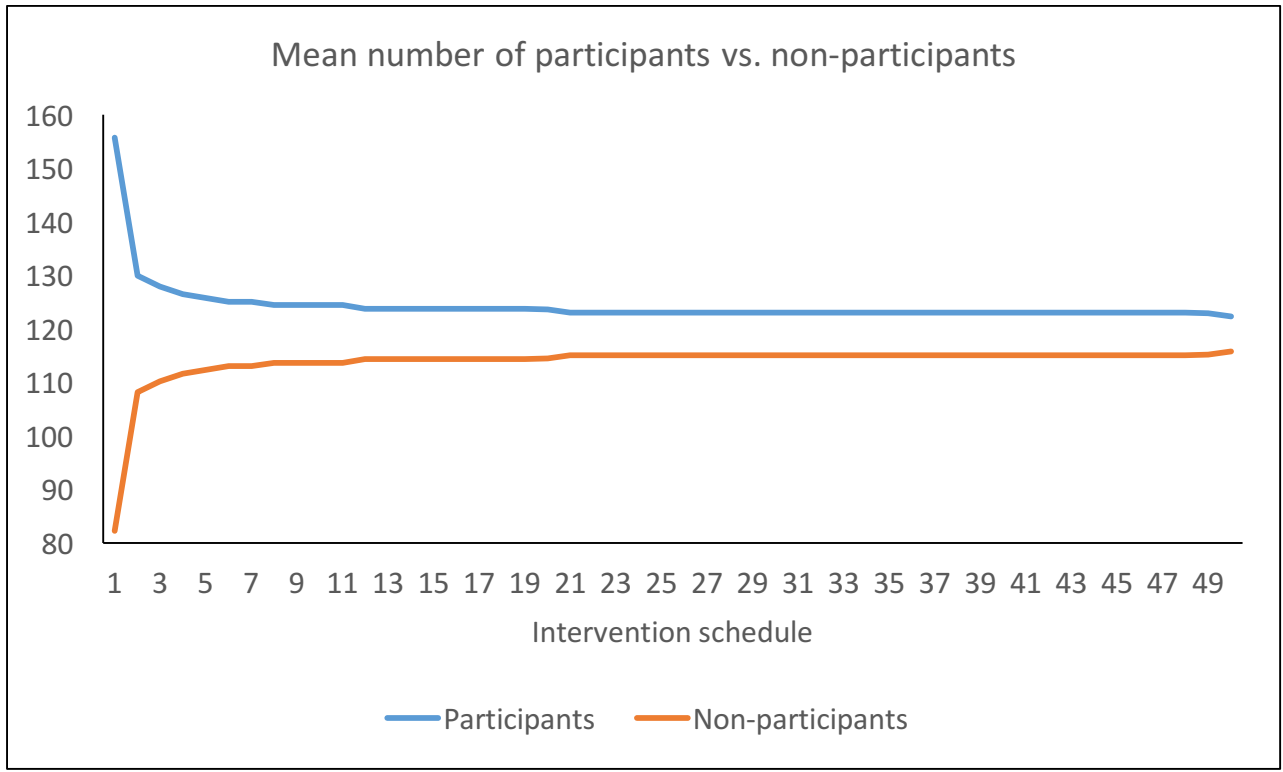

**Fig. 5** Change in mean number of participants and non-participants in simulation 2.



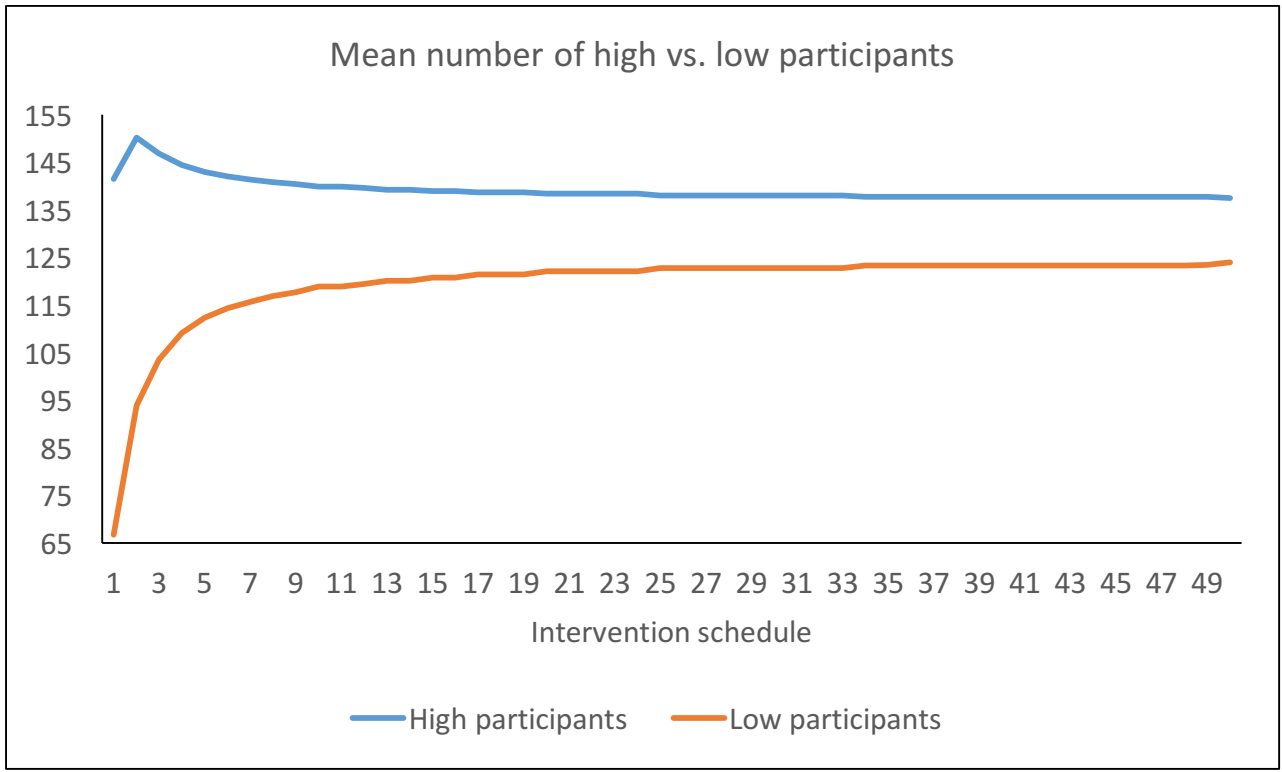

**Fig. 6** Change in mean number of high participants and low participants in simulation 3.



**Tables**

Table 1

*Sample ECM*

|            | A ($t$) | B ($t$) | C ($t$) |
|------------|---------|---------|---------|
| A ($t+1$)  | .70     | .50     | .10     |
| B ($t+1$)  | .20     | .30     | .20     |
| C ($t+1$)  | .10     | .20     | .70     |



Table 2

*ECM for three conditions* for Simulation 1 and 2

| Attainable moral story condition | Participant ($t$) | Non-participant ($t$) |
|---|---|---|
| Participant ($t+1$) | .90 | .44 |
| Non-participant ($t+1$) | .10 | .56 |
| **Extraordinary moral story condition** | | |
| Participant ($t+1$) | .64 | .12 |
| Non-participant ($t+1$) | .36 | .88 |
| **Control condition** | | |
| Participant ($t+1$) | .72 | .29 |
| Non-participant ($t+1$) | .28 | .71 |



Table 3

*Hypothetical ECM for three conditions* for Simulation 3 and 4

| Attainable story condition | High participant ($t$) | Low participant ($t$) | Non-participant ($t$) |
|---|---|---|---|
| High participant ($t+1$) | .30 | .20 | .10 |
| Low participant ($t+1$) | .60 | .70 | .60 |
| Non-participant ($t+1$) | .10 | .10 | .30 |
| Extraordinary story condition | | | |
| High participant ($t+1$) | .70 | .60 | .10 |
| Low participant ($t+1$) | .20 | .30 | .20 |
| Non-participant ($t+1$) | .10 | .10 | .70 |
| Control condition | | | |
| High participant ($t+1$) | .30 | .10 | .05 |
| Low participant ($t+1$) | .50 | .40 | .15 |
| Non-participant ($t+1$) | .20 | .50 | .80 |



Table 4

*Results of mixed-effects analyses for three different participant groups*

| | High participants | | | | Low participants | | | | Non-participants | | | |
|---|---|---|---|---|---|---|---|---|---|---|---|---|
| | *B* | *SE* | *z* | 95% CI | *B* | *SE* | *z* | 95% CI | *B* | *SE* | *z* | 95% CI |
| Schedule 1 | 648.20 | 37.90 | 17.10*** | [573. 92 722.48] | 589.23 | 50.13 | 11.75*** | [490.98 687.49] | -1237.43 | 28.78 | -43.00*** | [-1293.83 -1181.03] |
| Schedule 2 | 450.23 | 37.90 | 11.88*** | [375.95 524.51] | 509.45 | 50.13 | 10.16*** | [411.20 607.71] | -959.68 | 28.78 | -33.35*** | [-1016.09 -903.28] |
| Schedule 3 | 244.74 | 37.90 | 6.46*** | [170.45 319.02] | 967.88 | 50.13 | 19.31*** | [869.62 1066.14] | -1212.62 | 28.78 | -42.14*** | [-1269.02 -1156.21] |
| Schedule 4 | 764.96 | 37.90 | 20.18*** | [690.68 839.25] | 101.20 | 50.13 | 2.02* | [2.94 199.46] | -866.16 | 28.78 | -30.10*** | [-922.57 -809.76] |
| Wald $\chi^2(4)$ | 527.84*** | | | | 488.57*** | | | | 2454.44*** | | | |
| Snijders/Bosker $R^2$ | .45 | | | | .49 | | | | .68 | | | |
| Bryk/Raudenbush $R^2$ | .57 | | | | .49 | | | | .86 | | | |

*Note*. The schedule without any intervention application (schedule 5, control condition) is the reference group.

\* $p < .05$, \*\*\* $p < .001$



Table 5

*Results of ANOVA post-hoc tests for three different participant groups*

|  | Schedule 1 | | Schedule 2 | | Schedule 3 | | Schedule 4 | | Schedule 5 | |
|---|---|---|---|---|---|---|---|---|---|---|
|  | *M* | *SE* | *M* | *SE* | *M* | *SE* | *M* | *SE* | *M* | *SE* |
| High participants | 899.00[a] | 48.31 | 701.03[b] | 30.53 | 495.53[c] | 7.98 | 1015.76[a] | 35.78 | 250.80[d] | 2.14 |
| Low participants | 1310.73[a] | 47.01 | 1230.95[a] | 48.25 | 1689.38[b] | 37.84 | 822.70[c] | 19.05 | 721.50[c] | 3.72 |
| Non-participants | 790.27[a] | 29.98 | 1068.02[b] | 39.62 | 815.09[a] | 45.18 | 1161.54[b] | 17.24 | 2027.71[d] | 5.84 |

*Note.* Means in the same row with different subscripts are significantly different, $p < .05$ (Scheffe's method applied).